\documentclass{iopart}
\usepackage{graphicx}
\usepackage{amssymb}

\begin{document}

\title{Directional Locking and the Influence of Obstacle Density on Skyrmion Dynamics in Triangular and Honeycomb Arrays}
\author{N.P. Vizarim$^{1}$, J.C. Bellizotti Souza$^{2}$,  C. Reichhardt$^{3}$, C.J.O. Reichhardt$^{3}$, and P.A. Venegas$^{2}$}
\address{$^{1}$POSMAT - Programa de P{\' o}s-Gradua{\c c}{\~ a}o em Ci{\^ e}ncia e Tecnologia de Materiais, Faculdade de Ci\^{e}ncias, Universidade Estadual Paulista - UNESP, Bauru, SP, CP 473, 17033-360, Brazil}
\address{$^{2}$Departamento de F\'{i}sica, Faculdade de Ci\^{e}ncias, Universidade Estadual Paulista - UNESP, Bauru, SP, CP 473, 17033-360, Brazil
  }
\address{$^{3}$ Theoretical Division and Center for Nonlinear Studies,
Los Alamos National Laboratory, Los Alamos, New Mexico 87545, USA
}
\ead{cjrx@lanl.gov, nicolas.vizarim@unesp.br}

\begin{abstract}
  We numerically examine the dynamics of a single skyrmion driven over triangular and honeycomb obstacle arrays at zero temperature. The skyrmion Hall angle $\theta_{sk}$, defined as the angle between the applied external drive and the direction of the skyrmion motion, increases in quantized steps or continuously as a function of the applied drive. For the obstacle arrays studied in this work, the skyrmion exhibits two main directional locking angles of $\theta_{sk}=-30^\circ$ and $-60^\circ$. We show that these directions are privileged due to the obstacle landscape symmetry, and coincide with channels along which the skyrmion may move with few or no obstacle collisions. By changing the obstacle density, we can modify the skyrmion Hall angles and cause some dynamic phases to appear or grow while other phases vanish. This interesting behavior can be used to guide skyrmions along designated trajectories via regions with different obstacle densities. For fixed obstacle densities, we investigate the evolution of the locked $\theta_{sk}=-30^\circ$ and $-60^\circ$ phases as a function of the Magnus force, and discuss possibilities for switching between these phases using topological selection.
  
\end{abstract}

\maketitle

\vskip 2pc

\section{Introduction}

Overdamped particles driven over a periodic substrate can exhibit a series of 
directional locking steps of preferred motion. Depending on the substrate symmetry, the particles 
can flow in a direction which differs from that of the applied external drive.
For a square obstacle array,
locking motion appears along the angles
$\phi = \arctan(p/q)$, where $p$ and $q$ are integers.
The most notable locking steps appear at  $\phi=0^\circ$,
$45^\circ$ and $90^\circ$.
The locking steps differ for other obstacle array symmetries.
For example, in triangular 
arrays, the most prominent locking angles are
$\phi=30^\circ$ and $60^\circ$,
and locking motion appears at angles of
$\phi = \arctan(\sqrt{3}p/(2q+1))$.
Locking phenomena has been widely studied
using colloidal particles
\cite{Korda02,Gopinathan04,Balvin09,Reichhardt12,Stoop20,Reichhardt04b},
type II superconducting vortices
\cite{Reichhardt12,Reichhardt99,Silhanek03},
classical electrons \cite{Wiersig01}, and more 
recently using magnetic skyrmions
\cite{Vizarim20,Vizarim20d}
and active matter \cite{Reichhardt20}.
In the case of colloidal particles,
the colloids are placed within
a periodic array of posts or optical traps,
and an external drive is applied. Locking is observed as either the
driving force or the substrate is rotated.
The width of the locking steps can be modeled as
a function of the details of the colloid-substrate interaction such as
the shape and size of the obstacles or traps.
Directional locking also 
appears for quasiperiodic substrates,
which exhibit five or seven locking directions
\cite{Reichhardt11,Bohlein12a}.

Most directional locking studies have focused on overdamped systems,
and produce locking steps by changing the angle of an applied external
drive or by rotating the substrate.
For magnetic skyrmions, due to the Magnus force it is possible
to observe directional locking with the driving direction and substrate
orientation both fixed.
Skyrmions in chiral magnets can be found in 
numerous materials and their sizes can vary from a micron
to 10 nm \cite{Fert17}.
Recently, stable 
skyrmions were found in materials at room temperature
\cite{Woo16,Legrand17,Legrand20,Soumyanarayanan17}.
When skyrmions are 
subjected to an external drive they can be set into motion
and undergo a transition from a pinned to a sliding state 
\cite{Fert17,Schulz12,Jonietz10,Nagaosa13,Lin13}.
They can be measured in experiments by direct observation of skyrmion motion 
\cite{Woo16,Yu12,Montoya18}, changes in the topological Hall effect
\cite{Schulz12,Liang15}, x-ray diffraction \cite{Zhang18} or 
neutron scattering \cite{Okuyama19}.

Skyrmions are promising for many technological applications.
For example, skyrmions could be used
as bits for information storage in logical and memory devices
\cite{Kiselev11,Hagemeister15}  where the bit state is
associated with the presence or absence of a magnetic skyrmion.
The main difference between 
skyrmions and overdamped particles is the important role played by the
Magnus force, which may dominate the dynamics of a skyrmion system
\cite{OlsonReichhardt14}.
It is believed that the Magnus force is also the origin of the
low value of the observed threshold depinning current for skyrmions
\cite{Lin13a,Iwasaki13}. In the absence of 
defects in the sample, the skyrmion flows at
an angle with respect to the external drive
known as the intrinsic skyrmion Hall angle
$\theta_{sk}^{\rm int}$ \cite{Fert17,Nagaosa13,Jiang17}.
This angle changes as a function
of the ratio between the Magnus and the damping term.
Experimentally observed
skyrmion Hall angles
range from a few degrees up to values close to
90$^\circ$, but the higher angles may 
be achieved only in certain systems
\cite{Jiang17,Litzius17,Woo18,Juge19,Zeissler20}.
The intrinsic skyrmion Hall angle can be a 
problem for certain technological applications,
such as race track devices, since it limits 
the maximum distance a skyrmion may flow before
contacting the edge of the device. Thus, 
controlling the skyrmion motion is of
significant interest in the scientific community
\cite{Fert17}.
One way to achieve such control is
by introducing artificial pinning centers or obstacles to 
the sample.
The skyrmion Hall angle in the presence of pinning is close to 0$^\circ$ at
low drives,
but it can increase in quantized steps
as a function of the applied drive \cite{Vizarim20,Reichhardt15a}.
For sufficiently high drives, the skyrmion Hall angle saturates
to a value very close 
to the intrinsic Hall angle, $\theta_{sk}^{\rm int}$
\cite{Vizarim20,Jiang17,Reichhardt15a,Reichhardt16,Reichhardt18a,Kim17}.
Most studies of skyrmions
interacting with pinning involve
randomly placed defects that act like 
obstacles or pinning centers;
however, nanostructuring techniques
make it possible to place
the defects at specific positions,
enabling the creation of
ordered pinning arrangements \cite{Stosic17,Saha19}.

In this work we simulate the behavior of a single skyrmion
interacting with triangular 
and honeycomb obstacle arrays while being driven with an external force that is applied in the
$x$ direction. The system is at zero temperature and we study its behavior for varied 
values of the obstacle density and the ratio of
the Magnus to the damping term. We find that the 
skyrmion behavior is strongly dependent on the obstacle density,
and that some dynamical regimes appear or grow in magnitude while
others vanish.
For both arrays, the skyrmion exhibits 
strong directional locking at $\theta_{sk}=-30^\circ$ and $-60^\circ$ due to the
substrate symmetry.
We plot dynamical 
phase diagrams that summarize this strong locking effect and its stability.
Our results indicate that it is possible to use the obstacle 
density as a way to control the skyrmion motion.

\section{Simulation}

\begin{figure}
  \begin{center}
    \includegraphics[width=0.8\columnwidth]{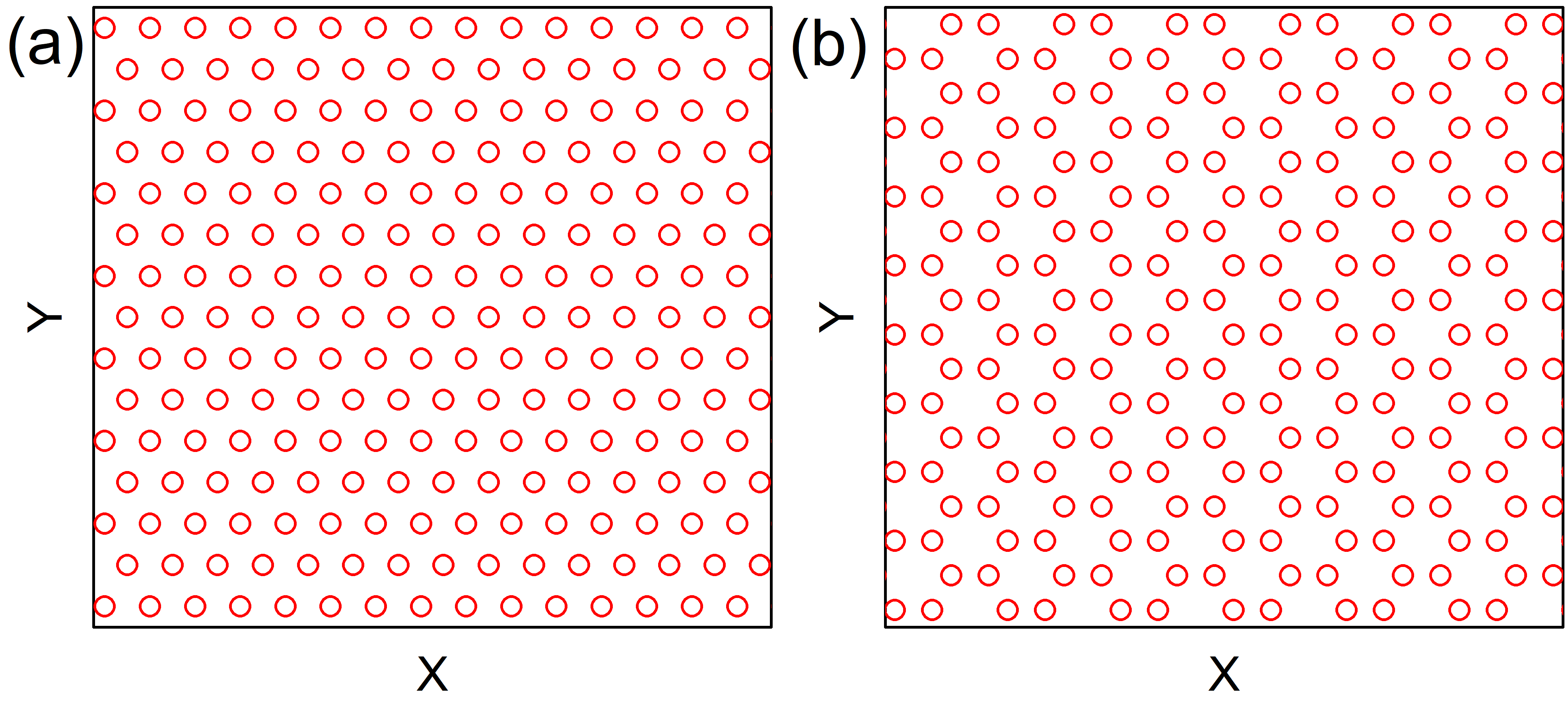}
    \end{center}
  \caption{Illustration of the obstacle arrays used in this work.
    (a) Triangular array. (b) Honeycomb array.
}
\label{fig:1}
\end{figure}

We consider the dynamics of a single skyrmion in a two-dimensional system
of size $L \times L$ with periodic boundary conditions
in the $x$ and $y$ directions.
The skyrmion interacts with either a triangular or a honeycomb obstacle
array, illustrated
in Fig.~\ref{fig:1}.
We obtain the skyrmion dynamics
using
a particle-based model for skyrmions \cite{Lin13},
shown in Eq. (1), using an in-house simulation code based 
on standard Molecular Dynamics techniques \cite{Allen87}.

\begin{equation}
 {\alpha }_d{\rm {\bf{v}}}_i+{\alpha }_m\hat{z}\times {\rm {\bf{v}}}_i={\rm {\bf{F}}}^o_i+{\rm{\bf{F}}}^D .
\end{equation}

In this equation, the first term on the left is the damping arising
from the spin precession and dissipation of 
electrons localized in the skyrmion core,
where ${\alpha }_d$ is the  
damping constant.
The second term on the left represents the Magnus 
force, where ${\alpha }_m$ is the Magnus constant. The Magnus term produces a force that is 
perpendicular to the skyrmion velocity.
The first term on the right of Eq. (1) is the 
interaction between the skyrmion and the obstacles. We model
this potential with the Gaussian 
form \cite{Vizarim20,Vizarim20d,Vizarim20a}
$U_o=C_oe^{-{\left({r_{io}}/{a_o}\right)}^2}$, where $C_o$ is the strength 
of the obstacle potential, $r_{io}$ is the distance between skyrmion $i$ and
obstacle 
$o$, and $a_o$ is the obstacle radius. Thus, the force between the obstacle
and the skyrmion 
takes the form
${\rm {\bf{F}}}^o_i=-\mathrm{\nabla }U_o=-F_or_{io}e^{-{\left({r_{io}}/{a_o}\right)}^2}{\widehat{\rm {\bf {r}}}}_{io}\ $, where $F_o=2C_o/a^2_o$.
Throughout this work we set $a_0=0.65$ and $F_0=1$. For computational 
efficiency we set a cut-off at $r_{io}=2.0$;
beyond this length the interaction is negligible.
All distances are normalized by
the screening length $\xi$, and
obstacle densities are given in terms of $1/\xi^2$.
The last term in Eq.~(1) is the external dc drive,
${\rm {\bf {F}}}^D =F^D\widehat{\rm {\bf{d}}}$,
where $\widehat{\rm {\bf{d}}}$  is the direction of the
applied external driving force. In 
this work we fix
$\widehat{\rm {\bf{d}}}=\ \widehat{\rm {\bf{x}}}$. We measure the skyrmion velocity parallel, $\left\langle V_{\parallel }\right\rangle $, and perpendicular, $\left\langle V_{\bot }\right\rangle $ to the drive.
For a skyrmion moving in the absence of obstacles, in the 
overdamped limit of ${\alpha }_m/{\alpha }_m =0$, the skyrmion
flows in the same direction as
the driving force. For finite values of ${\alpha }_m/{\alpha }_m$,
however, the skyrmion 
motion follows the skyrmion Hall angle,
$\theta_{sk}={\mathrm{arctan} \left(\left\langle V_{\bot }\right\rangle /\left\langle V_{\parallel }\right\rangle \right)\ }={\mathrm{arctan} \left({\alpha }_m/{\alpha }_d\right)\ }$. The external driving force is increased in small 
steps of $\delta F=0.001$ and
we wait ${10}^5$ simulation time steps between
drive increments to ensure
that the system has reached a steady state.
We normalize the damping and Magnus coefficients
according to ${\alpha }^2_d+{\alpha }^2_m=1$.

\section{Directional locking on triangular and honeycomb obstacle arrays} 

\begin{figure}
  \begin{center}
    \includegraphics[width=0.6\columnwidth]{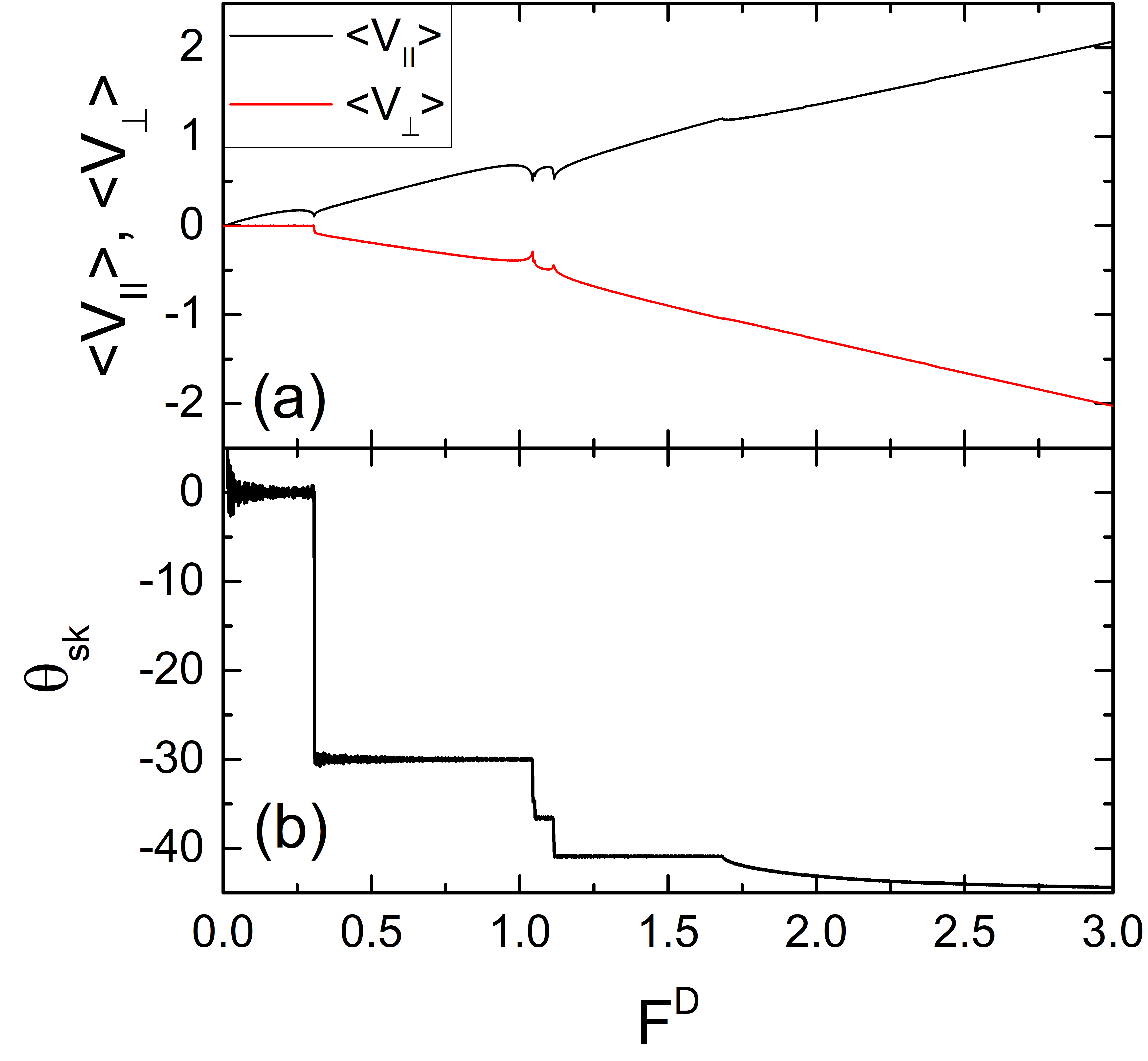}
    \end{center}
\caption{(a) $\langle V_{||}\rangle$ (black) and $\langle V_{\perp}\rangle$ (red)
  vs $F^{D}$ for a system containing a single skyrmion interacting with a triangular obstacle array with density  $\rho_t = 0.128$ and $\alpha_{m}/\alpha_{d} = 1.0$.
  (b) The corresponding skyrmion Hall angle $\theta_{sk}$ vs $F^{D}$.
} 
\label{fig:2}
\end{figure}

\begin{figure}
  \begin{center}
    \includegraphics[width=0.6\columnwidth]{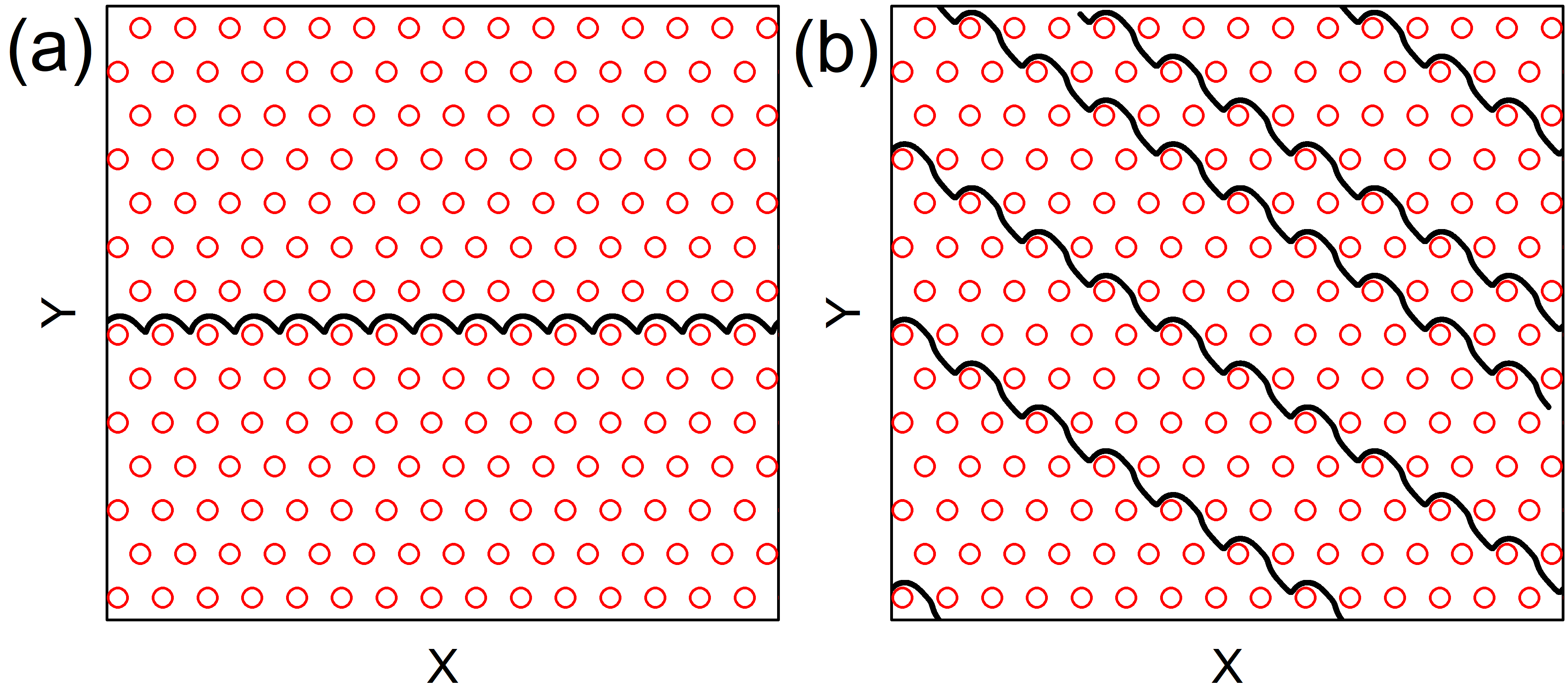}
    \includegraphics[width=0.6\columnwidth]{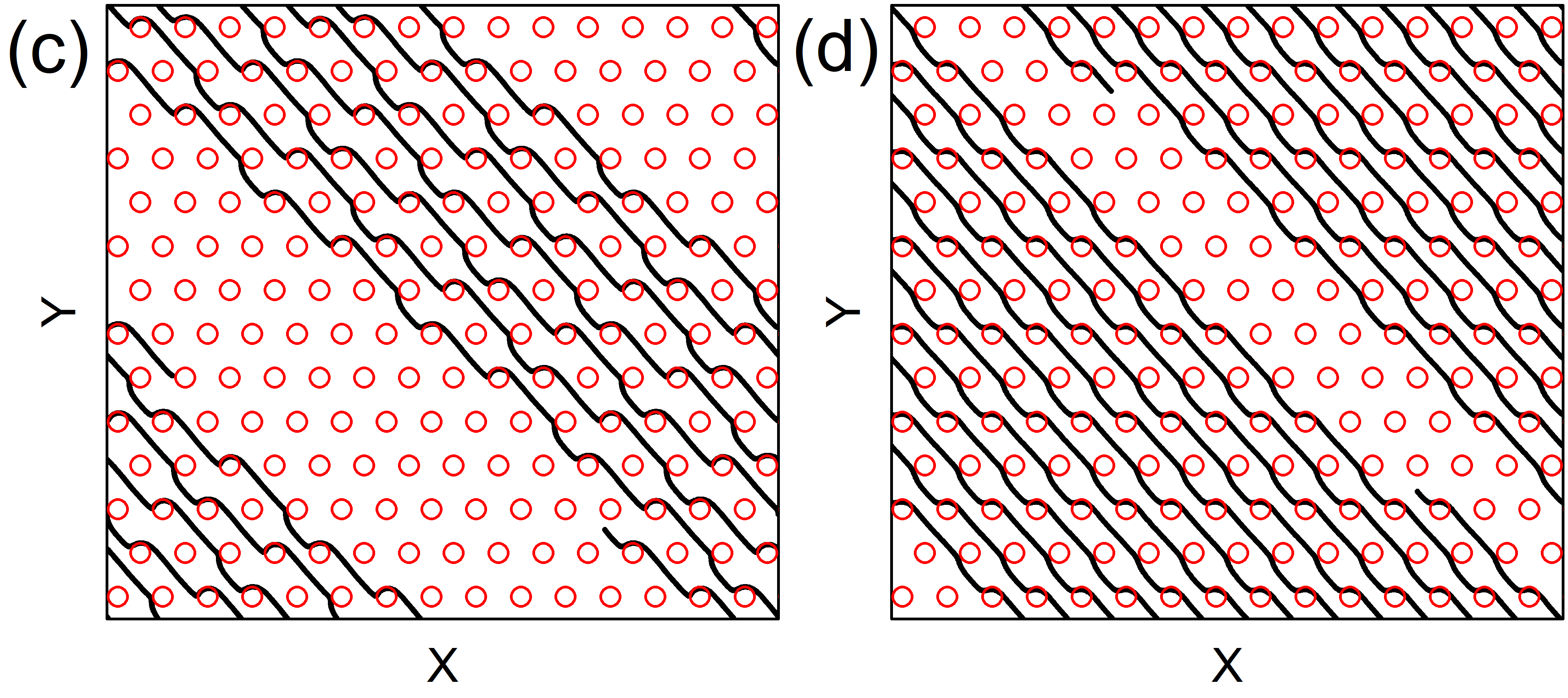}
    \end{center}
\caption{ Obstacles (open circles) and the skyrmion trajectory (black lines)
  for the triangular array system from Fig.~\ref{fig:2}
  with $\rho_t = 0.128$ and $\alpha_{m}/\alpha_{d} = 1.0$.
  (a) At $F^{D} = 0.25$, the skyrmion is oscillating
  in the $y$ direction due to the obstacle centers and flowing
  in the $+x$ direction with $\theta_{sk}=0^\circ$.
  (b) At $F^{D} = 0.5$, the skyrmion exhibits motion in both
  the $x$ and $y$ directions with $\theta_{sk}=-30^\circ$.
  (c) At $F^{D} = 1.1$, $\theta_{sk}=-36.56^\circ$.
  (d) At $F^{D} = 1.5$, $\theta_{sk}=-40.8^\circ$. }    
\label{fig:3}
\end{figure}

In Fig.~\ref{fig:2}(a) and (b) we show
$\langle V_{||}\rangle$, $\langle V_{\perp}\rangle$, and $\theta_{sk}$ as
a function of $F_D$ for a sample with a triangular
obstacle array where
$\rho_t = 0.128$ and $\alpha_{m}/\alpha_{d} = 1.0$.
In the absence of obstacles, the 
skyrmion flow would be along the intrinsic Hall angle,
$\theta_{sk}^{int}=-45^\circ$. 
As shown in Fig.~\ref{fig:2}(b),
when obstacles are present the skyrmion
direction of motion exhibits a series of quantized locking steps
up to $F^D=1.7$, followed at higher drives by a continuous saturation
of $\theta_{sk}$ to the intrinsic Hall angle.
For $F^D\leqslant0.307$ the
skyrmion motion is locked to
$\theta_{sk}=0^\circ$, as shown in Fig.~\ref{fig:3}(a),
and Fig.~\ref{fig:2}(a)
indicates that over this range of drives,
$\langle V_{||}\rangle$ is increasing with $F^D$ while
$\langle V_{\perp}\rangle = 0$.
For $0.307<F^D\leqslant1.041$ the skyrmion
locks to
$\theta_{sk}=-30^\circ$, which is a preferred direction of
motion due to the symmetry of the 
triangular obstacle array.
Here the skyrmion moves by one lattice constant in the $x$ direction for
every lattice constant in the $y$
direction,
giving
$\theta_{sk}=\arctan(\sqrt{3}p/(2q+1)) = \arctan(\sqrt{3}/3) = 30^\circ$.
The corresponding skyrmion trajectory in Fig.~\ref{fig:3}(b)
flows around the upper contour of one obstacle in each row before proceeding
to the next row.
In Fig.~\ref{fig:3}(c) we 
illustrate the motion for
locking to $\theta_{sk}=-36.56^\circ$ at $F^D=1.1$.
In this case the skyrmion 
trajectories are straighter due to the higher velocities.
At $F^D=1.5$ in
Fig.~\ref{fig:3}(d),
the skyrmion trajectories
are along $\theta_{sk}=-40.8^\circ$.
For higher values of the 
applied drive, the skyrmion
direction of motion continuously approaches the intrinsic skyrmion Hall angle.

\begin{figure}
  \begin{center}
    \includegraphics[width=0.6\columnwidth]{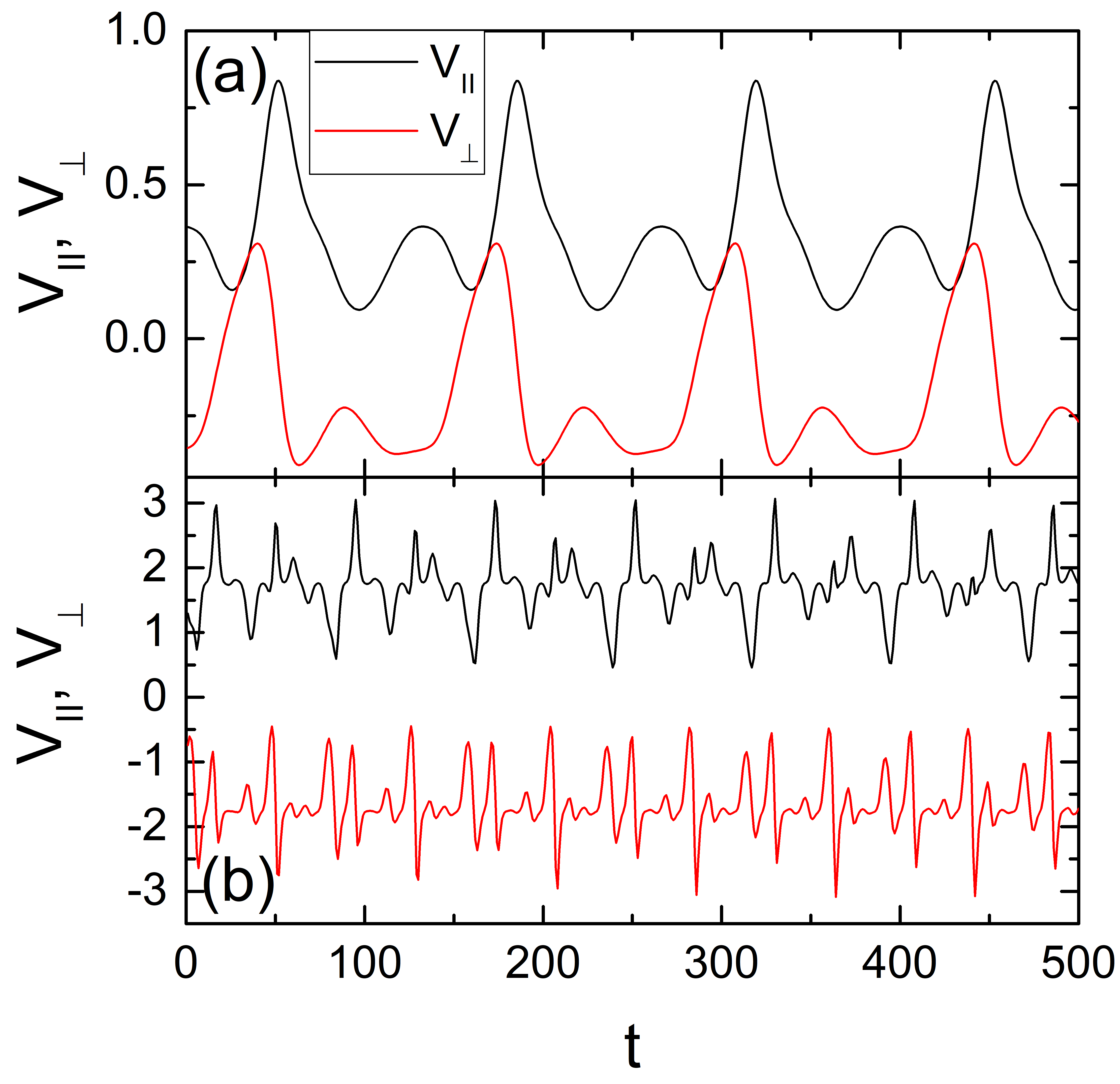}
    \end{center}
  \caption{ Time series of the velocities in the parallel
    direction, $V_{||}$ (black), and the perpendicular direction,
    $V_{\perp}$ (red) for the
    triangular array system in Fig.~\ref{fig:2} with $\rho_t=0.128$
    and $\alpha_m/\alpha_d=1.0$.
    (a) A locking regime
    at $F^{D} = 0.5$ with $\theta_{sk} = -30^\circ$.
    (b) A nonlocking regime at $F^{D} = 2.5$.
    In the locking regime, the signals repeat themselves exactly, while
    in the non-locking regime they show some variations.
    The variations are most clearly visible in $V_{||}$
    in panel (b), where there is a
    sequence of a tall peak followed by a group of shorter peaks; the heights
    of these shorter peaks vary from cycle to cycle.
  }    
\label{fig:4}
\end{figure}

\begin{figure}
  \begin{center}
    \includegraphics[width=0.6\columnwidth]{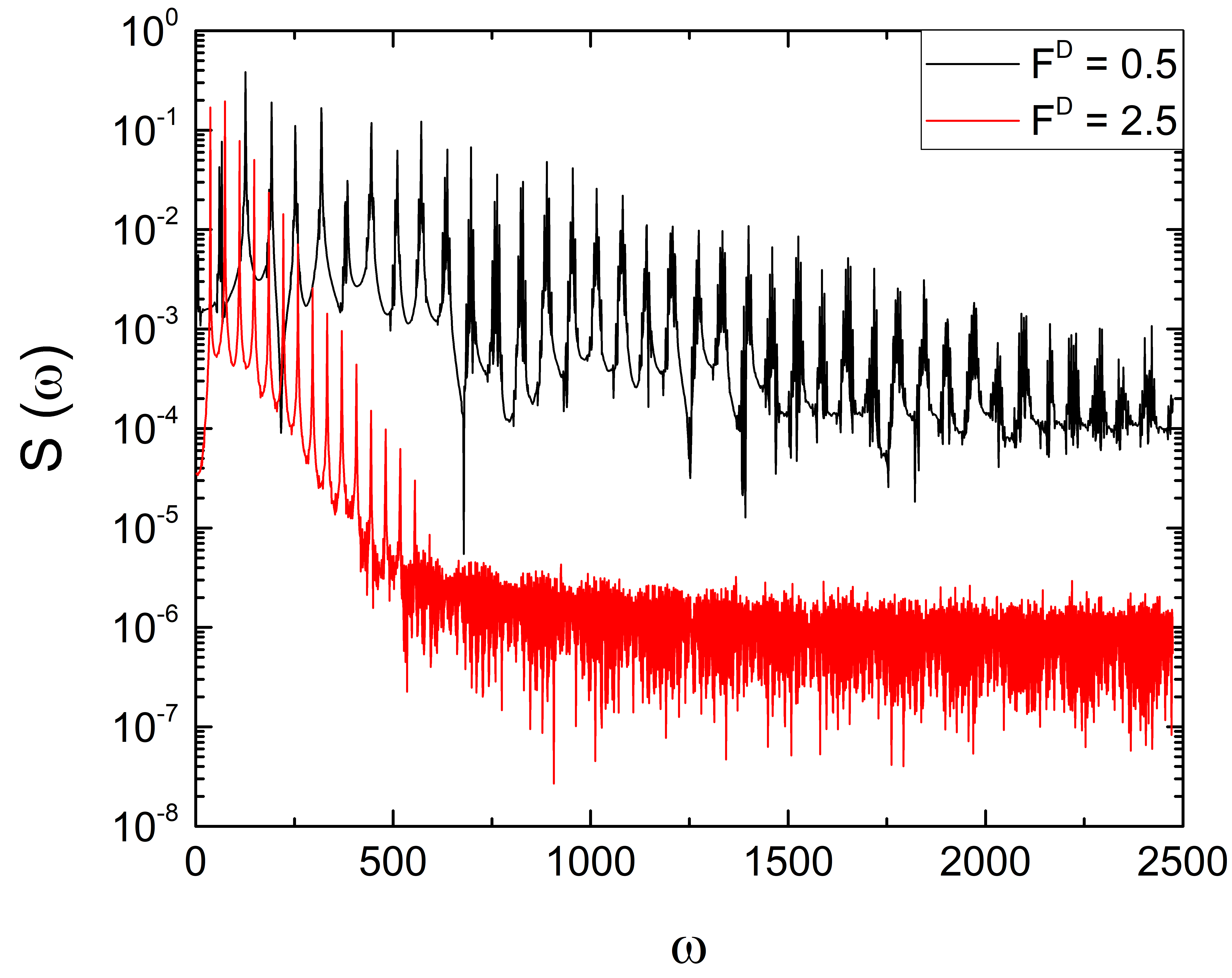}
    \end{center}
  \caption{The Fourier transform $S(\omega)$ of the $V_{||}$ data in
    Fig.~\ref{fig:4} for the
    triangular array system with $\rho_t=0.128$
    and $\alpha_m/\alpha_d=1.0$ in
    the locking regime at $F^D=0.5$ (black) and
    the non-locking regime at $F^D=2.5$ (red).
    The peaks are much stronger in the locking regime.
}    
\label{fig:5}
\end{figure}

In addition to producing steps, the locking
is also
visible in the 
time series of the particle velocities.
In Fig.~\ref{fig:4}(a) we show the parallel and
perpendicular velocities at
$F^{D} = 0.5$ when the system is in a directionally locked phase
with $\theta_{sk}=-30^{\circ}$,
while Fig.~\ref{fig:4}(b) shows the same
quantities at
$F^{D} = 2.5$ in a non-step region.
In the locked phase, the velocities are exactly periodic
with the same signal appearing during each cycle.
In the non-locked region, there are more 
oscillations for the
same time because the drive is higher; however,
the patterns are no longer 
exactly repeating, although there is still
a periodicity in the signal due to to the periodic array.
The differences can be seen 
more clearly
by examining the Fourier transform $S(\omega)$ of the
velocity signals for the $x$ direction velocity component $V_{||}$,
as shown in Fig.~\ref{fig:5}.
The peaks in $S(\omega)$ are much stronger in the locking regime than in
the non-locking regime.
This suggests that if an ac driving is added to the dc driving,
the locking step regions should
show strong Shapiro step phenomena whereas in
the non-locking regions, the Shapiro step effect will be absent or reduced.

\begin{figure}
  \begin{center}
    \includegraphics[width=0.6\columnwidth]{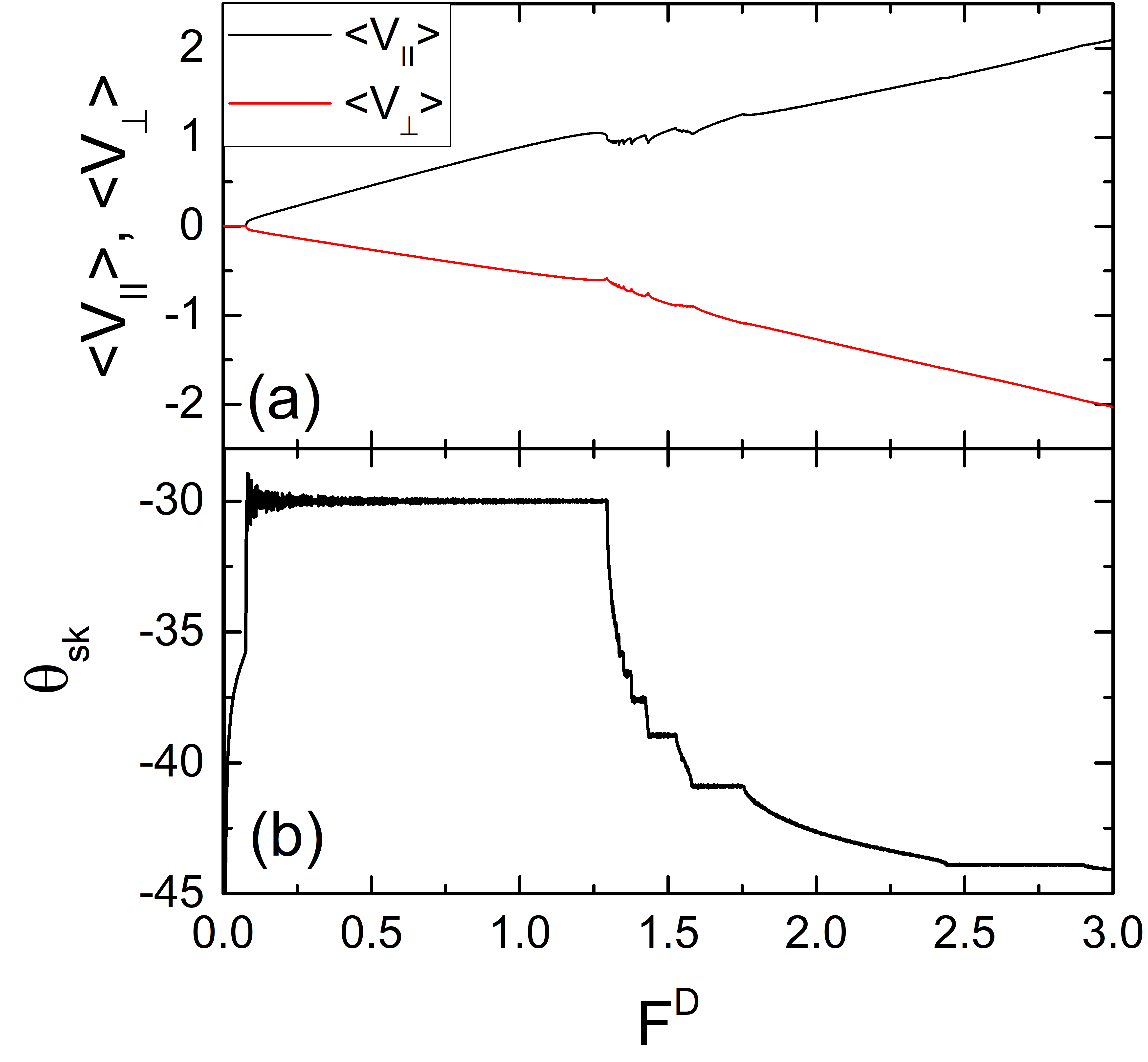}
    \end{center}
\caption{(a) $\langle V_{||}\rangle$ (black) and $\langle V_{\perp}\rangle$ (red)
  vs $F^{D}$ for a system containing a single skyrmion interacting with
  a honeycomb obstacle array with density  $\rho_t = 0.123$
  and $\alpha_{m}/\alpha_{d} = 1.0$.
  (b) The corresponding skyrmion Hall angle $\theta_{sk}$ vs $F^{D}$.}
\label{fig:6}
\end{figure}

\begin{figure}
  \begin{center}
    \includegraphics[width=0.6\columnwidth]{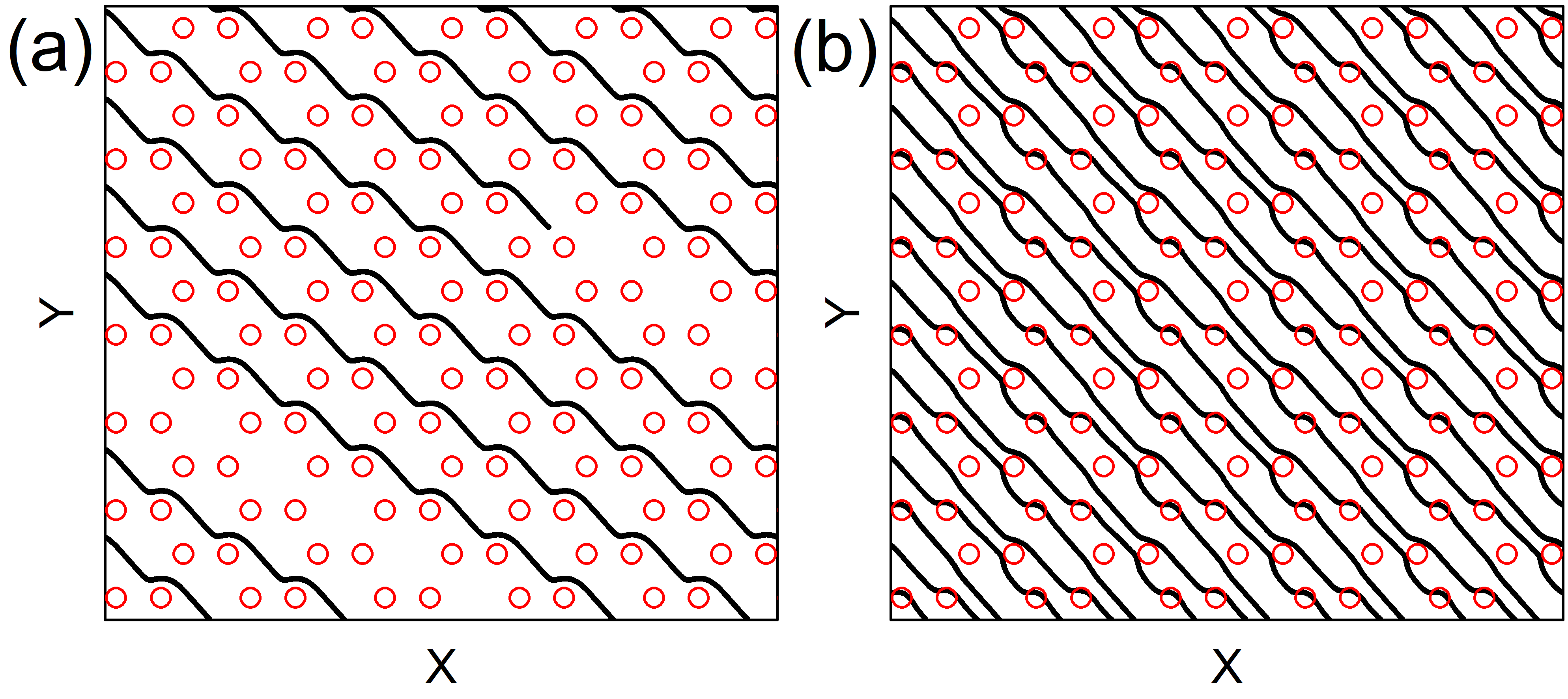}
    \end{center}
  \caption{Obstacles (open circles) and the skyrmion trajectory (black
    lines) for
    the honeycomb array system
    from Fig.~\ref{fig:6} with $\rho_h = 0.123$ and
    $\alpha_{m}/\alpha_{d} = 1.0$.
  (a) At $F^{D} = 0.5$, the skyrmion is flowing with $\theta_{sk}=-30^\circ$.
  (b) At $F^{D} = 1.7$, $\theta_{sk}=-40.8^\circ$.   
 }
\label{fig:7}
\end{figure}

Turning next to the honeycomb obstacle lattice,
in Fig.~\ref{fig:6} we plot $\langle V_{||}\rangle$, $\langle V_{\perp}\rangle$,
and $\theta_{sk}$ versus $F^D$
for a system with $\rho_h = 
0.123$ and $\alpha_{m}/\alpha_{d} = 1.0$.
There is
a pinned phase for $F^D\leqslant0.076$
when the skyrmion is trapped between the obstacles.
For higher $F^D$, the 
skyrmion depins and moves
along $\theta_{sk}=-30^\circ$, as illustrated in Fig.~\ref{fig:7}(a).
Note that unlike
the triangular array, there is no regime with
$\theta_{sk}=0^\circ$ since
the vacancies in the honeycomb lattice trap the skyrmion at low drives.
For $F^D>1.294$, there is a series of transitions in
the skyrmion motion, which ultimately saturates at a values
close to the intrinsic Hall angle
when $F^D=3.0$.
As an example of the motion found on these closely spaced short steps,
in Fig.~\ref{fig:7}(b) we show the
skyrmion trajectory at $F^D=1.7$ where 
$\theta_{sk}=-40.8^\circ$.

\section{Effect of Varied Obstacle Density and Magnus Force}

\begin{figure}
  \begin{center}
    \includegraphics[width=0.9\columnwidth]{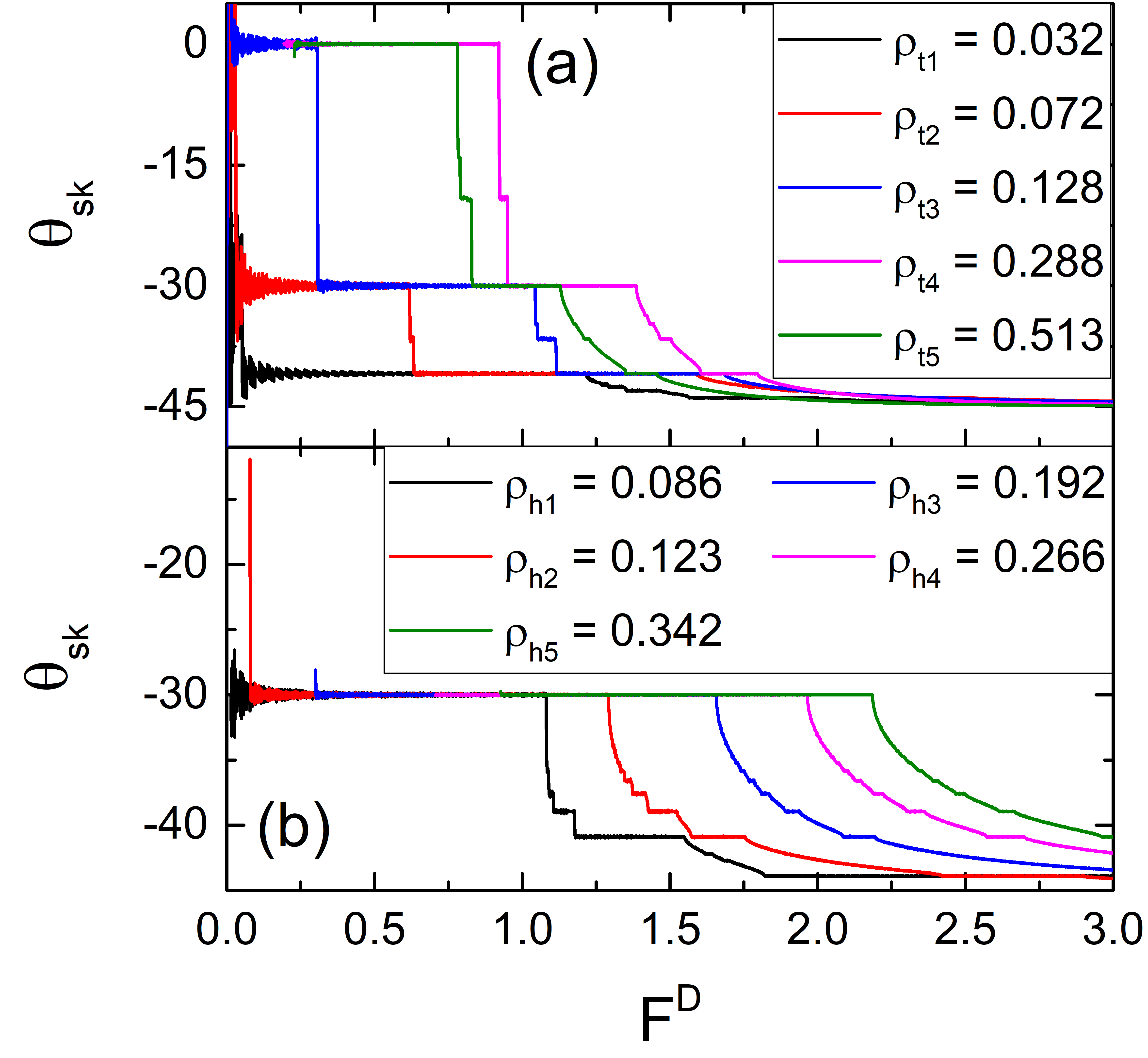}
    \end{center}
  \caption{ Plots of the skyrmion Hall angle $\theta_{sk}$ vs
    $F^D$ for several values of the
    obstacle density at $\alpha_{m}/\alpha_{d} = 1.0$.
    (a) A triangular obstacle array with 
    $\rho_t = 0.032$ (black), $\rho_t = 0.072$ (red), 
    $\rho_t = 0.128$ (blue), $\rho_t = 0.288$ (magenta),
    and $\rho_t = 0.513$ (green). 
    (b) A honeycomb obstacle array with
    $\rho_h =  0.086$ (black), $\rho_h = 0.123$ (red),
    $\rho_h = 0.192$ (blue), $\rho_h = 0.266$ (magenta),
    and $\rho_h = 0.342$ (green).
}
\label{fig:8}
\end{figure}

To study the influence of the obstacle density on
the skyrmion dynamics, we fix $\alpha_{m}/\alpha_{d}=1.0$ and
consider five samples
with different obstacle densities for
both the triangular and honeycomb arrays.
For each sample we then separately
vary $\alpha_{m}/\alpha_{d}$ in order to 
determine the influence of the Magnus force on the behavior.
In Fig.~\ref{fig:8}, we plot $\theta_{sk}$ versus $F^D$ for 
varied obstacle densities for the triangular and honeycomb arrays,
showing a clear influence of obstacle density on the skyrmion dynamics.
In Fig.~\ref{fig:8}(a), at very low values of obstacle density
such as $\rho_t = 0.032$, the skyrmion motion is
very close to the intrinsic skyrmion Hall angle, which for 
$\alpha_{m}/\alpha_{d}=1.0$ is $\theta_{sk}^{\rm int}=-45^\circ$.
That is, the skyrmion flow is nearly unaffected by the presence
of the widely spaced obstacles. 
As more obstacles 
are added to the sample, the skyrmion dynamics become richer and
multiple locking steps of different sizes emerge.
For the triangular array 
there are three main locking directions
of $0^\circ$, $-30^\circ$, and $-40.8^\circ$.
The $\theta_{sk}=0^\circ$ locking occurs only for 
higher obstacle densities of $\rho_t > 0.128$, while
the $\theta_{sk}=-30^\circ$ locking is 
present for all density values simulated in this work,
although its extent is limited to
$0\leqslant F^D\leqslant0.05$
when $\rho_t = 0.032$.
The
$\theta_{sk}=-40.8^\circ$
locking step is also present for all density values,
but it is more robust for low
densities.
In the honeycomb array, we observe only two main locking directions, 
$\theta_{sk}=-30^\circ$ and $-40.8^\circ$.
The $\theta_{sk}=-30^\circ$ locking is very 
robust for all density values,
but the phase $\theta_{sk}=-40.8^\circ$ is not very robust and becomes
weak at high obstacle densities.

\begin{figure}
  \begin{center}
    \includegraphics[width=0.6\columnwidth]{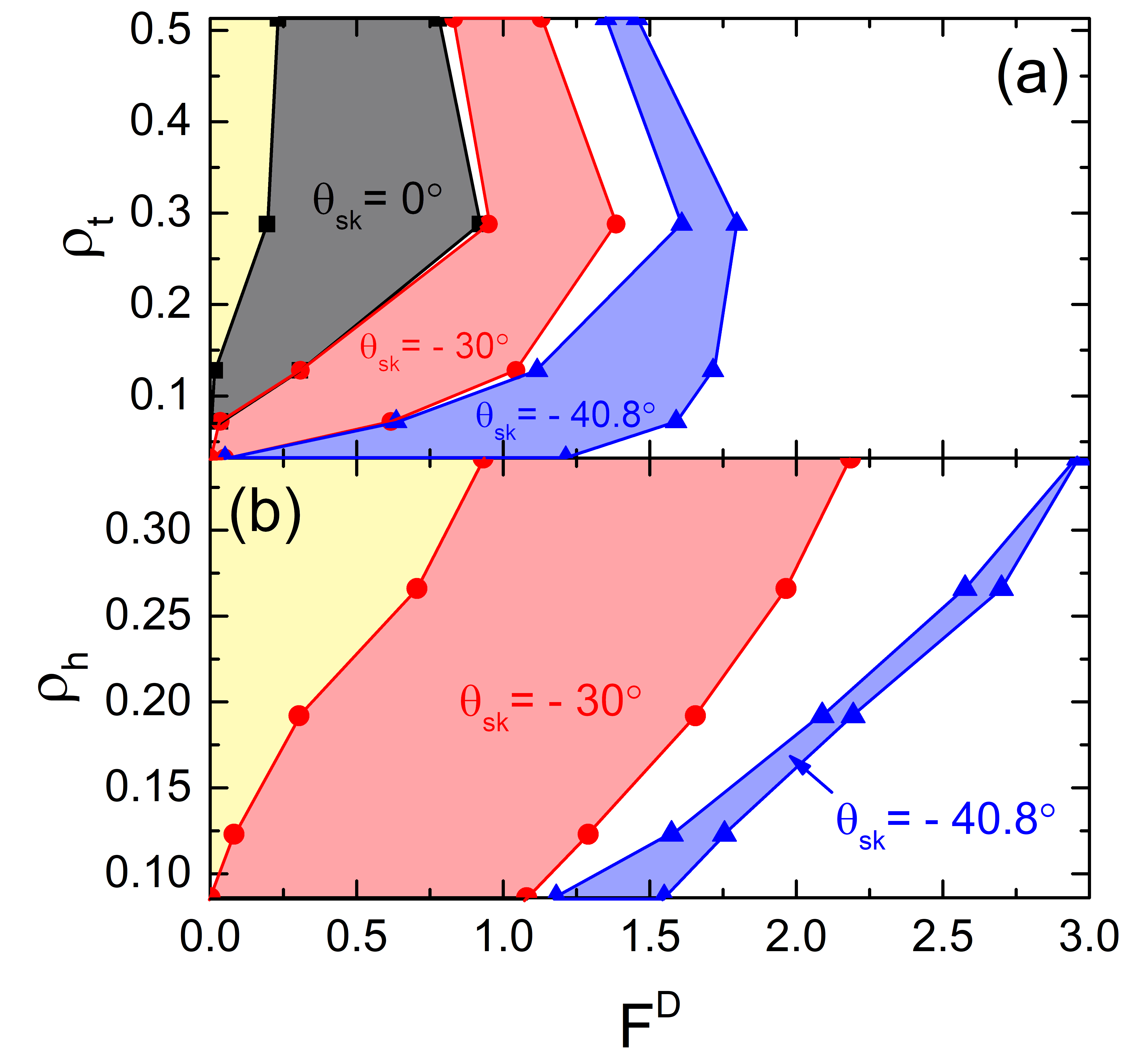}
    \end{center}
\caption{
Dynamic phase diagrams as a function of obstacle density vs $F^D$
for samples with $\alpha_{m}/\alpha_{d} = 1.0$. 
(a) The triangular obstacle array with obstacle density $\rho_t$.
(b) The honeycomb obstacle array with obstacle density $\rho_h$.
Colors indicate the different dynamical phases:
pinned (yellow), 
$\theta_{sk}=0^\circ$ (gray),
$\theta_{sk}=-30^\circ$ (red),
$\theta_{sk}=-40.8^\circ$ (blue),
and unlocked or minor locked states (blank areas).
}
\label{fig:9}
\end{figure}

We use the results obtained in Fig.~\ref{fig:8} to construct
dynamic phase diagrams of the main locking phases
as a function of obstacle density
$\rho_t$ or $\rho_h$ versus $F^D$, as shown
in Fig.~\ref{fig:9}.
In both cases, the extent of the pinned
phase increases as the obstacle density is increased.
The pinned phase is wider in the 
honeycomb lattice due to the trapping of the skyrmion inside the
lattice vacancy sites.
For the triangular lattice
in Fig.~\ref{fig:9}(a), the
$\theta_{sk}=0^\circ$ phase is absent for very low
obstacle densities and increases in width
up to a 
maximum at $\rho_t=0.288$.
In contrast, the
$\theta_{sk}=-40.8^\circ$ is largest
at low obstacle densities.
For the honeycomb lattice in Fig.~\ref{fig:9}(b),
the
$\theta_{sk}=-30^\circ$ step retains a nearly uniform width
over the entire range of obstacle densities considered, although the
location of the step shifts to higher $F_D$ with increasing $\rho_h$.
The $\theta_{sk}=-40.8^\circ$ is considerably diminished in extent
compared to
the triangular array.

\begin{figure}
  \begin{center}
        \includegraphics[width=0.9\columnwidth]{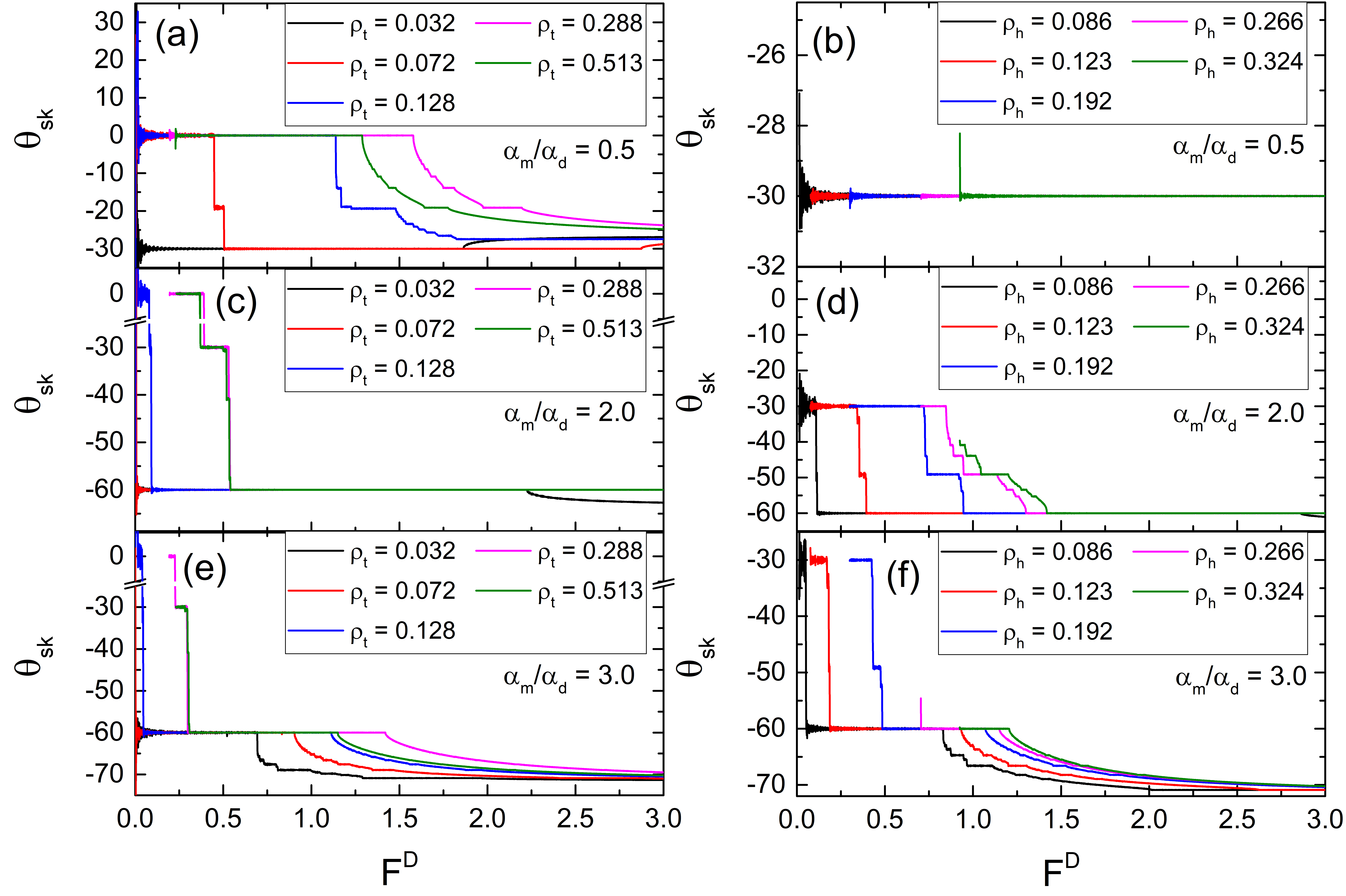}
    \end{center}
\caption{ 
  Plots of the skyrmion Hall angle $\theta_{sk}$ vs $F^D$ for
  different values of $\alpha_m/\alpha_d$ at varied obstacle densities.
  Left column: triangular obstacle arrays with $\rho_t=0.032$ (black),
  0.072 (red), 0.128 (blue), 0.288 (magenta), and 0.513 (green).
  Right column: honeycomb obstacle arrays with $\rho_h=0.086$ (black),
  0.123 (red), 0.192 (blue), 0.266 (magenta), and 0.324 (green).
  (a) and (b): $\alpha_m/\alpha_d=0.5$.
  (c) and (d): $\alpha_m/\alpha_d=2.0$.
  (e) and (f): $\alpha_m/\alpha_d=3.0$.
}
\label{fig:10}
\end{figure}

\begin{figure}
  \begin{center}
    \includegraphics[width=0.6\columnwidth]{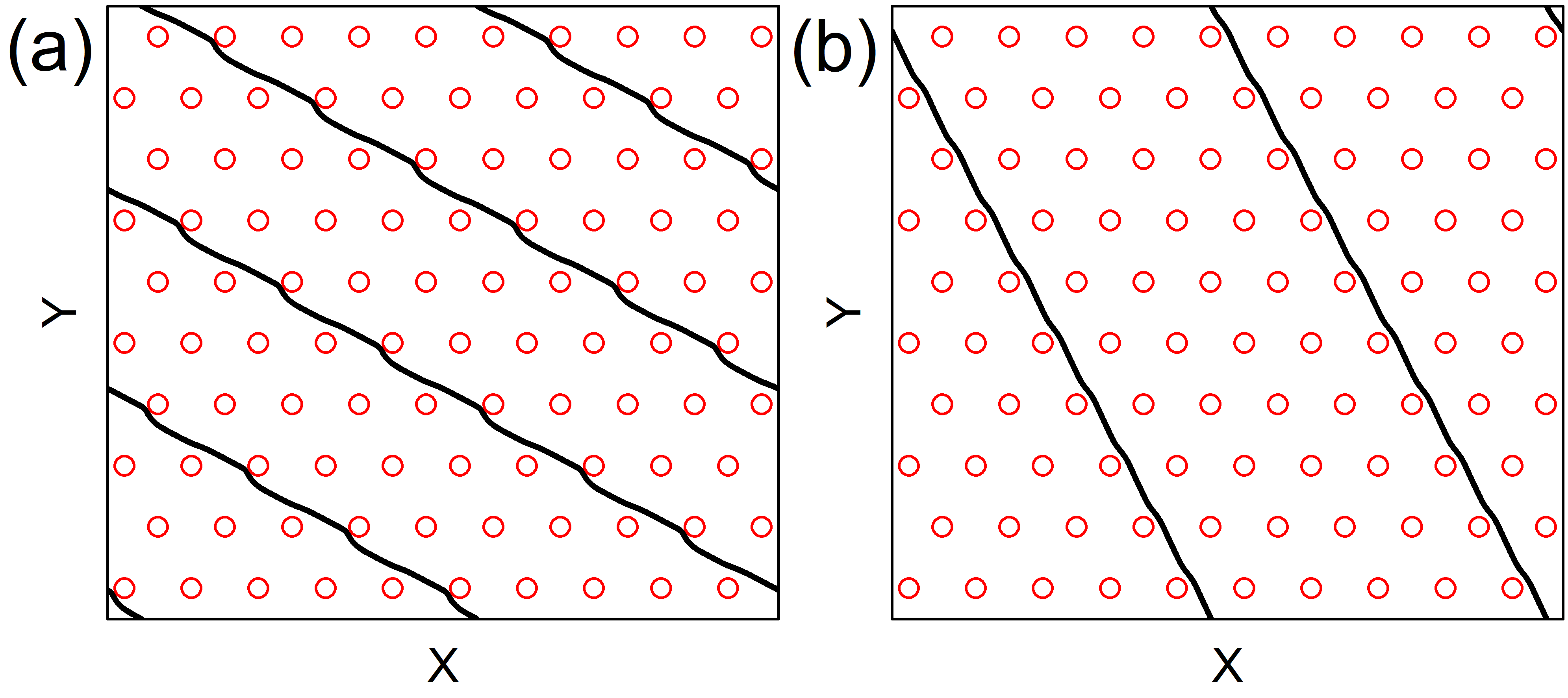}
    \includegraphics[width=0.6\columnwidth]{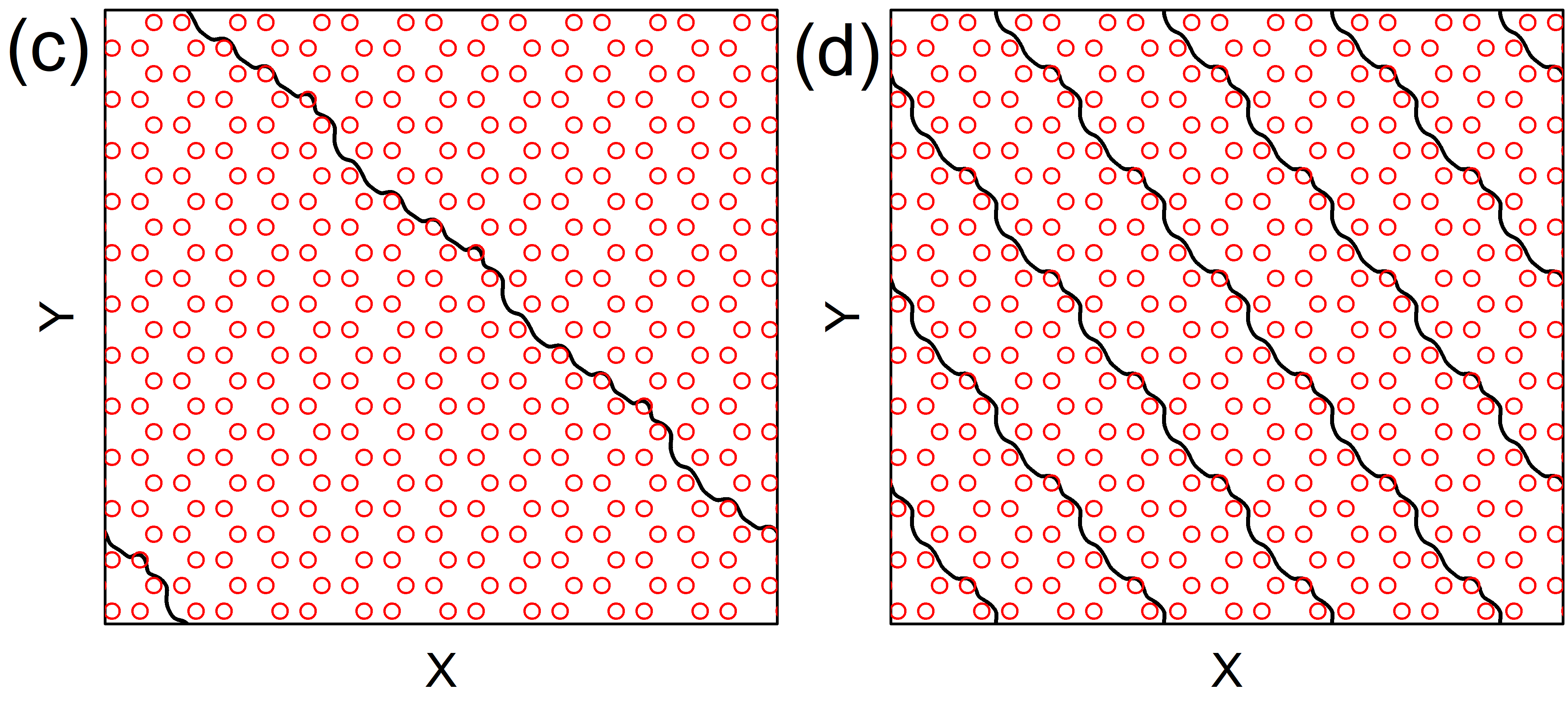}
    \end{center}
  \caption{Obstacles (open circles) and the skyrmion trajectory (black lines)
    for (a,b) a triangular obstacle array with $\rho_t=0.072$
    and (c,d) a honeycomb obstacle array with $\rho_h=0.324$.
    (a)
    At $F^D=1.5$ and
    $\alpha_{m}/\alpha_{d} = 0.5$, the motion is locked to
    $\theta_{sk}=-30^\circ$.
    (b)
    At $F^D=1.0$
    and $\alpha_{m}/\alpha_{d} = 2.0$,
    the motion is locked to $\theta_{sk}=-60^\circ$.
    (c)
    At $F^D=0.95$
    and $\alpha_{m}/\alpha_{d} = 2.0$,
    we find $\theta_{sk}=-40.8^\circ$.
    (d) At $F^D=1.125$
    and $\alpha_{m}/\alpha_{d} = 2.0$, 
 the motion is locked to $\theta_{sk}=-49^\circ$. 
}
\label{fig:11}
\end{figure}

In Fig.~\ref{fig:10} we plot $\theta_{sk}$ versus $F^D$ for different
values of $\alpha_m/\alpha_d$ at varied obstacle densities for both the
triangular and honeycomb arrays.
Note that the $\alpha_{m}/\alpha_{d}=1.0$ case is already 
shown in Fig.~\ref{fig:8}.
%
Figure~\ref{fig:10} indicates that the locking angles which appear
change as $\alpha_m/\alpha_d$ is modified.
When $\alpha_{m}/\alpha_{d}=0.5$, the intrinsic skyrmion Hall angle is 
$\theta_{sk}^{\rm int}=-26.57^\circ$.
As a result, the skyrmion can lock strongly to the
substrate symmetry direction of
$\theta_{sk}=-30^\circ$
even though $|\theta_{sk}| > |\theta_{sk}^{\rm int}|$ for this particular case.
This obstacle-induced motion at an angle with a magnitude larger than that
of the intrinsic skyrmion Hall angle
was observed in previous works for a
square obstacle array \cite{Vizarim20,Reichhardt15a}, but here we
show that
for a triangular obstacle array it depends upon the
obstacle density.
As illustrated in Fig.~\ref{fig:10}(a),
the $\theta_{sk}=-30^\circ$ locking step only exists for very 
low obstacle densities, such as $\rho_t<0.072$, and disappears at
higher obstacle densities.
In Fig.~\ref{fig:11}(a) we show the skyrmion trajectory for the
$\theta_{sk}=-30^\circ$ step at $F^D=1.5$, $\rho_t=0.072$, and
$\alpha_m/\alpha_d=0.5$.
In Fig.~\ref{fig:10}(a),
the number of dynamic phases which appear
is related to the obstacle density. At $\rho_t=0.128$, the system has 
a very rich set of dynamic phases, while for other densities the
number of dynamic phases is 
reduced.
For very low $\rho_t$ the skyrmion motion is
very close to the intrinsic Hall angle value,
so there are few dynamic phases.
On the other hand, when $\rho_t$ becomes too high,
the reduced spacing between adjacent obstacles pinches off many
possible skyrmion trajectories, eliminating the corresponding
dynamic phases.
%
We find that for a given value of
$F^D$,
varying the obstacle density can cause 
the skyrmion motion to follow a variety of different directions.
For example, at $F^D=0.75$, the skyrmion can move along
$\theta_{sk}=0^\circ$ or $\theta_{sk}=-30^\circ$ 
depending on the obstacle density. For 
$\rho_t\leqslant0.072$, the skyrmion locks to $\theta_{sk}=-30^\circ$, while for 
$\rho_t>0.072$, it locks to $\theta_{sk}=0^\circ$.
This opens the possibility of designing 
devices in which regions with distinct obstacle densities coexist
in order to force the skyrmion to follow a designated trajectory.
The pinning geometry is important, however, as
Fig.~\ref{fig:10}(b) indicates that in a honeycomb obstacle array,
there is only a single dynamic state of motion along
$\theta_{sk}=-30^\circ$ regardless of the obstacle density.

At $\alpha_{m}/\alpha_{d}=2.0$,
shown in Fig.~\ref{fig:10}(c,d), $\theta_{sk}^{\rm int}=-63.44^\circ$.
For the triangular array in Fig.~\ref{fig:10}(c),
when $\rho_t\leqslant0.072$ the skyrmion depins
directly into the 
$\theta_{sk}=-60^\circ$ step, illustrated in Fig.~\ref{fig:11}(b).
This is a very stable angle of motion for the 
triangular array due to its symmetry.
When $\rho_t\geqslant0.128$, additional steps emerge at
$\theta_{sk}=0^\circ$ and $-30^\circ$.
For the honeycomb lattice in Fig.~\ref{fig:10}(d), when
$\rho_h\leqslant0.266$
the skyrmion depins into the
$\theta_{sk}=-30^\circ$ state and follows the symmetry direction of the honeycomb 
lattice.
In contrast, at
$\rho_h=0.342$ the skyrmion depins
into the $\theta_{sk}=-40.8^\circ$ state shown in
Fig.~\ref{fig:11}(c).
The number of dynamic phases increases as the obstacle density increases.
For example, the
$\theta_{sk}=-49^\circ$ step illustrated in Fig.~\ref{fig:11}(d)
is absent for $\rho_h\leqslant0.123$.
In Fig.~\ref{fig:10}(e,f) we show samples with
$\alpha_{m}/\alpha_{d}=3.0$, where the intrinsic Hall angle is 
$\theta_{sk}^{\rm int}=-71.57^\circ$.
For each array, the locking steps at
$\theta_{sk}=-30^\circ$ and $-60^\circ$ are significantly reduced
in size. As the magnitude of the Magnus term
increases, the skyrmion moves at a larger angle with respect to the driving
direction,
reducing the
robustness of locking steps along
the preferred symmetry directions of the obstacle arrays.

\section{Stability of Directional Locking as a Function of $\alpha_{m}/\alpha_{d}$}

\begin{figure}
  \begin{center}
    \includegraphics[width=0.6\columnwidth]{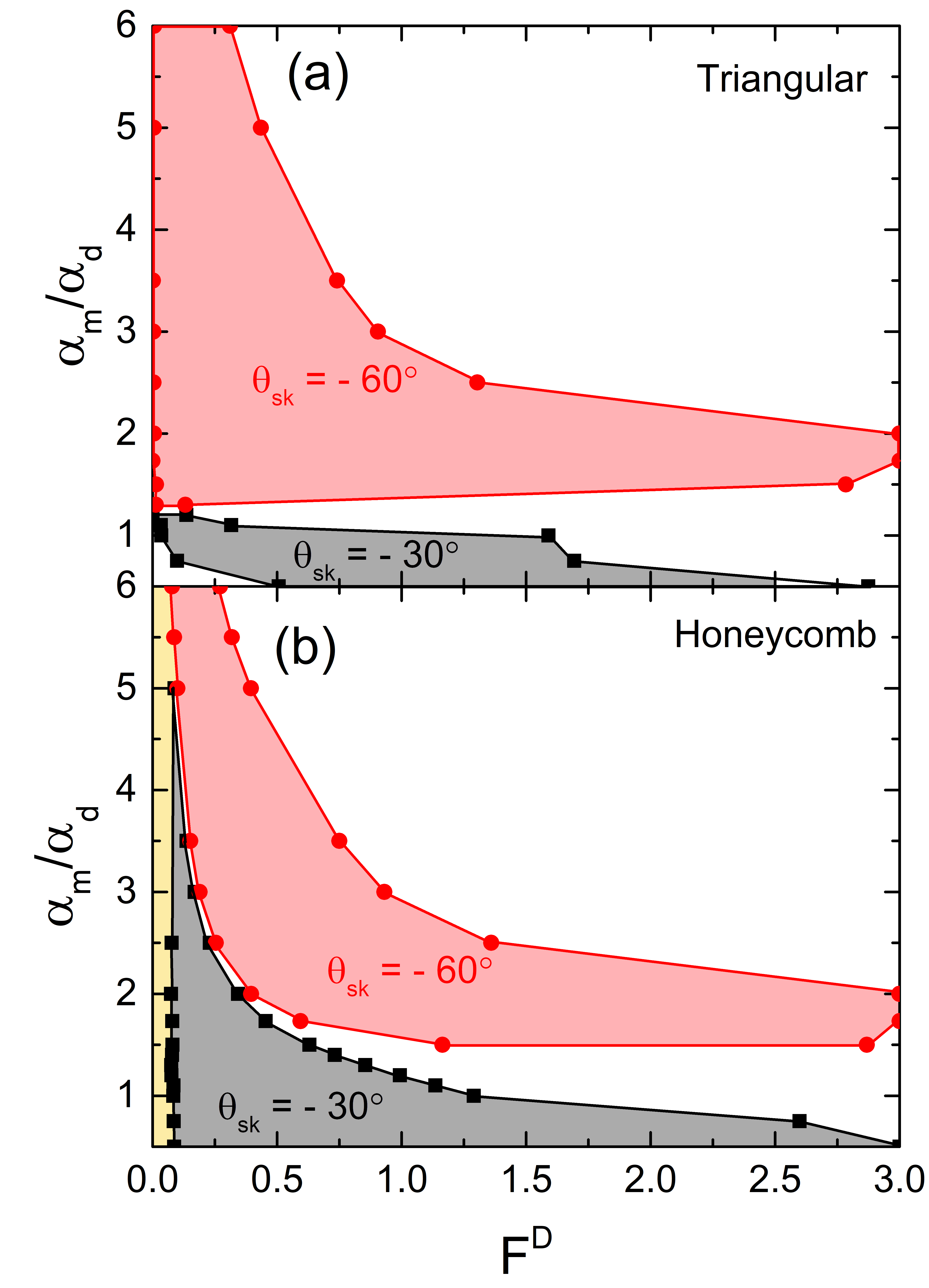}
  \end{center}
  \caption{Dynamic phase diagrams as a function of
    $\alpha_{m}/\alpha_{d}$ vs $F^D$.
    (a) The triangular obstacle array at $\rho_t = 0.072$.
    (b) The honeycomb obstacle array at $\rho_h = 0.123$.
    The different regions are:
    $\theta_{sk}=-30^{\circ}$ (black);
    $\theta_{sk}=-60^{\circ}$ (red);
    pinned (yellow);
    minor steps or nonstep regions (blank areas).
}
\label{fig:12}
\end{figure}

Both the triangular and honeycomb obstacle arrays should have preferred
directions of motion along
$\theta_{sk}=-30^\circ$ and $-60^\circ$ due to the array symmetry.
In this section we
quantify
the robustness of these dynamic phases as 
a function of $\alpha_{m}/\alpha_{d}$ for arbitrary obstacle densities.
We also
explore the
possibility of topological selection,
since different species of skyrmions with distinct 
Magnus forces may coexist in a given sample
\cite{Legrand17,Soumyanarayanan17,Jiang15,Nayak17,Karube18,Kovalev18,Ritzmann18}.
Our results are for the dynamics of a single skyrmion, but we
expect them to remain valid
for a system with 
a low density of skyrmions \cite{Vizarim20}.
We fix the obstacle density for the triangular 
$(\rho_t = 0.072)$ and honeycomb $(\rho_h = 0.123)$ arrays and
vary $\alpha_{m}/\alpha_{d}$ to 
investigate how the $\theta_{sk}=-30^\circ$ and $-60^\circ$
steps evolve.
In Fig.~\ref{fig:12} we plot
dynamic phase diagrams as a function of $\alpha_{m}/\alpha_{d}$
versus $F^D$ highlighting these selected locking steps.
For the triangular array in Fig.~\ref{fig:12}(a),
both steps have a range of $\alpha_m/\alpha_d$ for which the step
extends over nearly the entire window of $F_D$. This range is wider
for the $\theta_{sk}=-60^{\circ}$ step than for the
$\theta_{sk}=-30^{\circ}$ step.
This could be of interest for technological applications since the 
direction of the skyrmion motion remains unchanged on each step.
We find that when
$\alpha_{m}/\alpha_{d} \leqslant 1.2$,
the skyrmion only locks at $\theta_{sk}=-30^\circ$, while
for $\alpha_{m}/\alpha_{d} \geqslant 1.3$,
the skyrmion only locks with $\theta_{sk}=-60^\circ$.
That is,
the $\theta_{sk}=-30^\circ$ and $-60^\circ$ steps do not coexist
for a given value of
$\alpha_{m}/\alpha_{d}$.
This feature could be employed to perform
topological selection, since 
skyrmions with a stronger Magnus component could
lock to a different direction than
skyrmions with weaker Magnus components.
In the honeycomb array, Fig.~\ref{fig:12}(b) shows that
the $\theta_{sk}=-30^\circ$ and $-60^\circ$
steps coexist over the range 
$1.5 \leqslant \alpha_{m}/\alpha_{d} \leqslant 5.0$.
Here, the skyrmion can be switched from
$\theta_{sk}=-30^\circ$ motion to
$\theta_{sk}=-60^\circ$ motion by a fine adjustment of the external force.
For 
$\alpha_{m}/\alpha_{d} < 1.5$ the skyrmion can only lock to
$\theta_{sk}=-30^\circ$, while 
for $\alpha_{m}/\alpha_{d} > 5.0$,
the skyrmion locks only to $\theta_{sk}=-60^\circ$.
As 
mentioned in the previous section,
the obstacle density plays an important role in determining the skyrmion 
dynamics, so we expect that changing the obstacle density may
result in different phase 
diagrams. 

\section{Discussion}

Using triangular or honeycomb obstacle arrays,
the skyrmion motion can be locked into a range of different directions, with
the strongest locking occurring for
$\theta_{sk}=-30^\circ$ and $-60^\circ$.
Both of these locking directions are robust against
changing the obstacle density or
the value of $\alpha_{m}/\alpha_{d}$.
For a given value of
the external applied drive,
different obstacle densities can produce different directions of
skyrmion motion.
This property could be harnessed
to build devices
containing 
regions with different obstacle densities which
can steer the skyrmion along a desired trajectory.
We expect that similar results would appear for other periodic array
geometries but that the angles of motion would differ
depending on the substrate symmetry.
The impact of thermal fluctuations would be an interesting issue to
address in a future study.
It is known that temperature 
can modify phase transition points, or
even cause them to vanish \cite{Vizarim20a}.
Thermal creep could also become relevant
\cite{Reichhardt18a}.
If attractive pinning sites are used instead 
of repulsive obstacle sites, we expect very similar results, but
have found that attractive pinning 
sites produce less pronounced locking steps \cite{Vizarim20}.
Our results are based on a
point-like model for skyrmions \cite{Lin13};
however, actual skyrmions have internal modes
that can be excited which can modify the skyrmion dynamics.
Additionally, the skyrmions may be distorted in shape by the 
driving force or
through interactions with the obstacles \cite{Iwasaki13}.
Such effects
could be
explored using continuum-based simulations. 
Although we consider only a single skyrmion,
we expect our results to be general for the case of
multiple skyrmions at sufficiently low density, so that skyrmion-skyrmion
interactions remain unimportant.
If the skyrmion density is higher,
it is possible that new dynamic phases will arise due to the 
collective behavior \cite{Reichhardt18}.
Lattice commensuration effects \cite{Reichhardt17} could also become important.
When the skyrmion
lattice is commensurate with the pinning or obstacle array, the
skyrmions can arrange themselves in such a way that
the skyrmion-skyrmion interactions cancel out, producing dynamics that
are similar to those found in
the single skyrmion case.

\section{Summary}
We have investigated the dynamics of a single skyrmion driven over triangular and honeycomb obstacle arrays under zero temperature in order to determine the
effects of changing the obstacle density and develop possible new ways to
control the skyrmion motion.
We show that the skyrmion exhibits a series of directional locking effects that can be quantized or continuous as a function of the applied drive.
For low obstacle densities, the depinning forces are very weak and the skyrmion tends to move very close to the intrinsic Hall angle, reducing the number of dynamic phases.
At higher obstacle densities,
the depinning force is larger
and
a richer variety of directional locking phases
appear.
The main difference between the triangular and honeycomb obstacle
arrays is the absence
of a $\theta_{sk}=0^\circ$ locking step in the honeycomb array.
The vacancies in the honeycomb lattice trap the skyrmions more effectively
and prevent them from moving until the drive is too large to permit motion
along the $\theta_{sk}=0^\circ$ direction.
For the triangular array,
$\theta_{sk}=0^\circ$ steps are more
prominent at higher obstacle densities and lower
values of $\alpha_{m}/\alpha_{d}$.
Both arrays show pronounced locking at $\theta_{sk}=-30^\circ$
and $-60^\circ$ due to the array symmetry.
These locking steps appear over
a wide range of obstacle densities and $\alpha_{m}/\alpha_{d}$ values.
We demonstrated the robustness of these phases
for a variety of $\alpha_{m}/\alpha_{d}$ values,
and discussed possibilities for switching the skyrmion motion
between the locking steps or using the locking effects to perform
topological selection of different skyrmion species. 

\ack
This work was supported by the US Department of Energy through
the Los Alamos National Laboratory.  Los Alamos National Laboratory is
operated by Triad National Security, LLC, for the National Nuclear Security
Administration of the U. S. Department of Energy (Contract No. 892333218NCA000001).
N.P.V. acknoledges
funding from
Funda\c{c}\~{a}o de Amparo \`{a} Pesquisa do Estado de S\~{a}o Paulo - FAPESP (Grant 2017/20976-3).

\section*{References}
\bibliographystyle{iopart-num}
\bibliography{mybib}
\end{document}